\documentclass[sigconf,nonacm]{acmart}

\usepackage{amsmath}
{ 
	\theoremstyle{plain}
	\newtheorem{assumption}{Assumption}
}
\usepackage{graphicx}

\usepackage{xcolor}

\usepackage{booktabs}

\DeclareMathOperator{\E}{{\mathbb{E}}}

\usepackage{graphics}
\usepackage{float}

\begin{document}

\title{Network experimentation at scale}

\author{Brian Karrer}
\affiliation{Facebook}
\email{briankarrer@fb.com}

\author{Liang Shi}
\affiliation{Facebook}
\email{liangshi@fb.com}

\author{Monica Bhole}
\affiliation{Facebook}
\email{mbhole@fb.com}

\author{Matt Goldman}
\affiliation{Facebook}

\author{Tyrone Palmer}
\affiliation{Facebook}

\author{Charlie Gelman}
\affiliation{Facebook}

\author{Mikael Konutgan}
\affiliation{Facebook}

\author{Feng Sun}
\affiliation{Facebook}

\begin{abstract}
We describe our framework, deployed at Facebook, that accounts for interference between experimental units through cluster-randomized experiments.  We document this system, including the design and estimation procedures, and detail insights we have gained from the many experiments that have used this system at scale.  We introduce a cluster-based regression adjustment that substantially improves precision for estimating global treatment effects as well as testing for interference as part of our estimation procedure. With this regression adjustment, we find that imbalanced clusters can better account for interference than balanced clusters without sacrificing accuracy. In addition, we show how logging exposure to a treatment can be used for additional variance reduction. Interference is a widely acknowledged issue with online field experiments, yet there is less evidence from real-world experiments demonstrating interference in online settings. We fill this gap by describing two case studies that capture significant network effects and highlight the value of this experimentation framework.
\end{abstract}

\maketitle

\renewcommand{\shortauthors}{Karrer et al.}

\section{Introduction}

Experimentation is ubiquitous in online services such as Facebook, LinkedIn~\cite{Ya2015}, Netflix~\cite{diamantopoulos2019engineering}, etc., where the effects of product changes are explicitly tested and analyzed in randomized trials.  Interference, sometimes referred to as network effects in the context of online social networks, is a threat to the validity of these randomized trials as the presence of interference violates the \textit{stable unit treatment value assumption} (SUTVA, see e.g.~\cite{rubin1974estimating, cox1958planning, Rubin1980}) important to the analysis of these experiments.  Colloquially, interference means that an experimental unit's response to an intervention depends not just on their own treatment, but also other units' treatments.  For example, consider a food delivery marketplace that tests a treatment which causes users to order deliveries faster. This could reduce the supply of delivery drivers to users in the control group, leading the experimenter to overstate the effects of the treatment. 

In this paper, we introduce a practical framework for running experiments that accounts for interference at scale and has been used for many large cluster-randomized trials at Facebook.  Because the intent of these experiments is accounting for network effects, a less technical concept than interference, we refer to the framework as running \textit{network experiments}.  The framework ensures that network experiments are as easy to setup and run as any other experiment at Facebook.  Our claim is not that network experiments entirely solve the problem of interference, but instead that they can be useful, scalable, general purpose tools to produce better estimates of global average treatment effects and quantify network effects.  As is well known for cluster-randomized trials~\cite{hayes2017cluster}, clustering reduces statistical power, and the bias reduction can be irrelevant compared to the increased noise.  To counter this, we provide Monte-Carlo power analysis and a new cluster-based agnostic regression adjusted estimator, and 
the framework defaults to running side-by-side unit-randomized and cluster-randomized trials which we refer to as a \textit{mixed experiment}, an imbalanced variation of the experimental design described by~\cite{Saveski2017}.  Mixed experiments enable estimating the presence and magnitude of interference, and can retroactively indicate whether clustering was helpful or not.

The details of our approach and insights from running network experiments at scale can enlighten future development and methodological research.  Past literature often focuses on experiments and estimation methods with size-balanced clusters (e.g. \cite{Rolnick2019, pouget2019testing, Saveski2017, Huan2015}), sometimes claimed to provide better statistical power. Beyond balance being a difficult restriction to enforce in practice, we find that imbalanced clusters combined with our agnostic regression adjusted estimator can better capture interference with similar, or even superior, statistical power than balanced clusters.  Agnostic regression adjustment for unit-randomized trials is described in~\cite{lin2013}, and for cluster-randomized trials in~\cite{Middleton2015} but only for size-balanced clusters due to concerns about estimator bias.  We show this bias is vanishing for online experiments that utilize a large number of clusters, and the variance reduction from our regression adjusted estimator, which differs from those in previous work and does not assume clusters of equal sizes, is substantial.  Additionally, we derive how logging which units receive treatment, so-called trigger logging~\cite{Ya2015}, can be leveraged in network experiments for even more variance reduction.

While there are many theoretical investigations of interference in online social networks, relatively few experiments are described in the literature, and they generally contain weak evidence of interference, perhaps due to imprecise estimation.  By running many network experiments, we have found a number of experiments with apparent and substantive SUTVA violations and we describe two experiments in detail, demonstrating the usefulness and flexibility of our framework.  To summarize, we provide:

\textbf{Systems contributions:}
\begin{itemize}
\item Framework for deploying network experiments at scale
\item Procedure for designing and maintaining clusters
\end{itemize}

\textbf{Methodological contributions:}
\begin{itemize}
\item Detailed approach to leverage trigger logging in analysis
\item Agnostic regression adjusted estimator suitable for network experiments with imbalanced clusters and achieving significant variance reduction
\end{itemize}

\textbf{Empirical contributions:}
\begin{itemize}
\item Cluster evaluation indicating that imbalanced clusters are often superior in terms of bias-variance tradeoffs
\item Analysis of two real-world experiments demonstrating substantial network effects
\end{itemize}

We will first review necessary background and provide further motivation in Section~\ref{sec:background}.  Then we discuss our framework's implementation in Section~\ref{sec:implementation}, and provide our methodological contributions in Section~\ref{sec:analysis}.  We elucidate our procedure and claims about designing clusters for network experiments in Section~\ref{sec:clusters}, and then apply our framework to two real-world case studies in Section~\ref{sec:examples}.   

\section{Background and motivation}
\label{sec:background}
To make our discussion concrete, we introduce notation where random variables are capital letters and vectors are bold.  Let $Y_i$ be the random outcome for a specific experimental unit $i$ and $W_i \in \{0, 1\}$ be the random binary treatment condition of unit $i$.  Ideally an experiment would estimate the population mean behavior
\begin{eqnarray}
\mu(\textbf{w}) = \E[Y_u | \textbf{W}=\textbf{w}]
\label{eq:pop_mean}
\end{eqnarray}
for an arbitrary vector of population treatment assignments $\textbf{w}$, where the expectation corresponds to randomly selecting a unit $u$ in the population.  In particular, the global average treatment effect given by $\tau = \mu(\textbf{1}) - \mu(\textbf{0})$, and related contrasts, are of primary interest.  In words, the ideal experiment provides an understanding of \textit{what would happen to a random unit when all units are placed in one condition versus another condition}.  SUTVA implies that $\mu(\textbf{w}) = \E[Y_u | W_u = w_u]$, and hence $\tau = \E[Y_u | W_u = 1] - \E[Y_u | W_u = 0]$.  Fundamentally, SUTVA implies symmetries in the function $\mu$, and these symmetries are leveraged to estimate $\tau$ from experiments. 

A typical unit-randomized test (AB test) that treats a random fraction $p$ of the units and ignores interference estimates
\begin{eqnarray}
\tau_{unit}(p) = \E[Y_u | W_u = 1, p] - \E[Y_u | W_u = 0, p],
\end{eqnarray}
where the presence of $p$ reminds us that this expectation also includes randomizing this fraction of other units into treatment.  This only equals $\tau$ if SUTVA holds, and can be quite different otherwise.

Depending on the treatment that is tested, as well as the outcomes of interest, SUTVA may not hold and there is interference present between units. Cluster-randomized trials are one approach to better account for interference among experimental units.  A cluster-randomized trial is distinguished from an AB test through randomizing clusters of units to conditions.  Assume the population is partitioned into clusters where $c_u$ is the cluster assignment for unit $u$.  If a random fraction $p$ of clusters are treated, then a typical cluster-randomized test would estimate
\begin{eqnarray}
\tau_{cluster}(p) = \E[Y_u | \textbf{W}_{c_u} = \textbf{1}, p] - \E[Y_u | \textbf{W}_{c_u} = \textbf{0}, p]
\end{eqnarray}
where the notation $\textbf{W}_{c_u}$ means the vector of treatment assignments restricted to cluster $c_u$.  This estimand may have less bias compared to the true $\tau$ than $\tau_{unit}$ since the random unit is placed into an experience ``closer" to the counterfactual situation of interest~\cite{Eckles2014}.  In order for $\tau_{cluster}$ to equal $\tau$, we require no interference between clusters and can have arbitrary interference within clusters.  This situation is referred to as \textit{partial interference}~\cite{Hudgens2008} and should be viewed as SUTVA for clusters.

Sometimes relevant clusters of experimental units that obey partial interference are evident.  For examples outside online experimentation, a developmental economics experiment might randomize remote villages to loan policies, a public health experiment might randomize hospitals to different hand-washing procedures, and an education experiment might randomize schools to alternative curriculum (see e.g.~\cite{Duflo2008}).  Within online experimentation, meaningful clusters could be a group of advertising campaigns run by the same advertiser or a group of accounts across different online services all owned by the same person.

Relevant clusters that capture interference can also be unclear.  When spatial interactions are suspected as the cause of interference, geographic clusters, such as zip codes or Google's GeoCUTS~\cite{Rolnick2019} might be appropriate.   Within an online social network, interactions between users are crucial to the user experience and often a suspected cause of interference.  These interactions can be logged and viewed as a weighted graph correlated with interference.  Clustering this graph and randomizing the resulting clusters has been proposed as graph-cluster randomization~\cite{Ugander2013,Saveski2017}.

Interference can also occur \textit{intentionally} for new product features producing shared experiences.  For example, a pair of users may both be required to have access to a new product feature to utilize it together.  A small user-randomized experiment would produce few such pairs, which may vastly limit the estimated value of the new feature.  Randomizing pairs of users to access the feature is an alternative, but a typical user in a small pair-randomized test would still observe the feature in a state of limited utility.  Randomizing access to a feature at the cluster-level can provide a more consistent user experience within clusters and an experience closer to the launch of the new feature.  

The degree to which clusters manage to capture relevant interference is an empirical question that can be confronted with experimental data.   A variety of hypothesis tests have been devised for the presence of interference or network effects, with or without clusters.  The general idea is that a lack of interference implies equality across different estimands, which each can be estimated from the same experiment.   On the analysis side, exposure modeling~\cite{aronow2017} utilizes importance sampling through reweighing results to estimate alternative estimands.  Another approach splits experimental units into focal and other units and estimates whether variation in the other units' treatments affects the focal units~\cite{Athey2018}.  On the design side, \cite{Saveski2017} advocate an experimental design with balanced clusters that runs a side-by-side unit and cluster-randomized trial, where significant differences between estimates provided by cluster-randomized and unit-randomized units indicates the presence of interference.  An imbalanced variant was used by~\cite{holtz2020reducing}, and our framework provides another imbalanced version of this mixed experimental design by default.

\begin{figure*}[h]
    \includegraphics[width=0.75\textwidth]{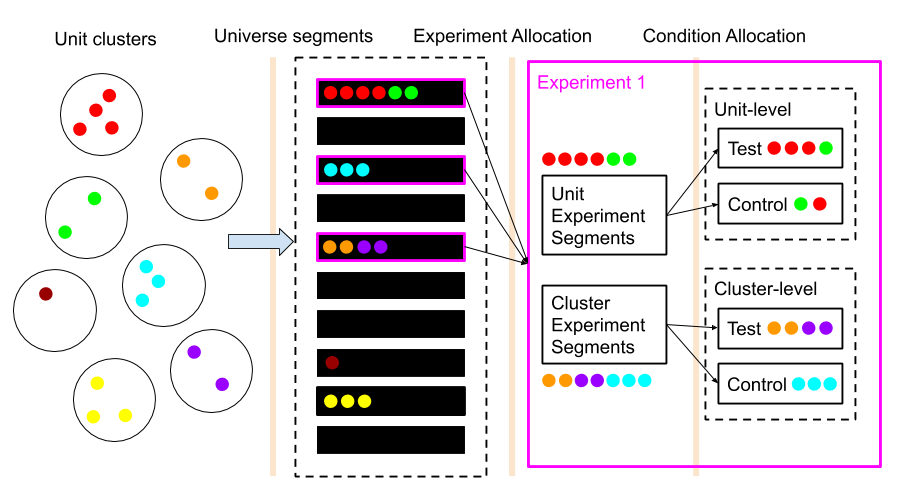}
    \caption{Visualization of the network experiment randomization process.}
    \label{fig:randomization}
\end{figure*}

\section{Network experiment implementation}
\label{sec:implementation}

Our implementation has two primary components that deploy treatments and manage clusterings respectively.  The component that deploys treatments is depicted visually in Figure~\ref{fig:randomization}, where the figure should be read from left to right.  A clustering of experimental units, represented by larger circles encompassing colored dots for units, is taken as input.  A given clustering and the associated units are considered as a \textit{universe}, the population under consideration.    

These clusters of experimental units are deterministically hashed into \textit{universe segments} based on the universe name, which are then allocated to experiments.  Universe segments allow a universe to contain multiple mutually exclusive experiments at any given time, a requirement for a production system used by engineering teams.  After allocation to an experiment, segments are randomly split via a deterministic hash based on the experiment name into unit-randomized segments and cluster-randomized segments.  The final condition allocation deterministically hashes units or clusters into treatment conditions, depending on whether the segment has been allocated to unit or cluster-randomization.  The result of this final hash produces the $\textbf{W}$ treatment vector that is used for the experiment.  For convenience, we denote $R_i = 1$ if unit $i$ was cluster-randomized and $R_i = 0$ if unit $i$ was unit-randomized.

\subsection{Trigger logging}
\label{subsec:trigger}

In practice, the allocation $W_i$ for a unit $i$ is only computed upon a call to the service supporting network experiments.  This call includes the universe name and experiment name, and a distributed key-value store is queried for the appropriate cluster of unit $i$ for that experiment.  If unit $i$ does not have a cluster or their cluster is not assigned to the experiment, they are excluded from the experiment's population, which in most cases means assigned to the control treatment.  Units in the experiment that call the service are logged along with their treatment assignment, which we follow~\cite{Ya2015} in referring to as \textit{triggering} their assignment. 

Trigger logging implies important invariances for network experiments that we will later leverage in analysis.  We let $T_i = 1$ if unit $i$ has a treatment assignment logged by the network experiment service (i.e. the unit is triggered) and $T_i = 0$ otherwise.  A unit's observed outcome cannot causally depend on their own treatment assignment unless they are triggered, as this treatment assignment is never used to do anything for that unit.  Hence $\E[Y_i | \textbf{W}_{-i}, W_i = x, T_i = 0]$ and $P(T_i = 0 | \textbf{W}_{-i}, W_i = x)$ do not depend on $x$, where $\textbf{W}_{-i}$ means the vector of treatment assignments excluding unit $i$.  However whether a unit is triggered can certainly depend on other units' treatment assignments, including units within their cluster, when interference exists.  An analogous statement in terms of a cluster as a whole can be phrased.  Let the notation $\textbf{T}_{c_i} = \textbf{0}$ mean that no unit in the cluster of unit $i$ is triggered.  Then we have that 
$\E[Y_i | \textbf{W}_{-c_i}, \textbf{W}_{c_i} = \textbf{x}, \textbf{T}_{c_i} = \textbf{0}]$ and $P(\textbf{T}_{c_i} = \textbf{0} | \textbf{W}_{-c_i}, \textbf{W}_{c_i} = \textbf{x})$ do not depend on $\textbf{x}$.

\subsection{Cluster management}

The other main component of the network experiment implementation is cluster management.  Clusterings can become stale or decrease in quality over time.  New unclustered units can arrive to the online service, causing a divergence between the experiment's population and the online service's population.  In addition, the pattern of interference can change over time, as might be detected by an increase in logged interactions between clusters.  To account for this, the framework provides two options: create a new universe or refresh the clustering used by an existing universe.

To accomplish either of these options, we have provided standardized pipelines that periodically regenerate clusterings or load in clusterings to our system.  A particular set of clusters is indexed both by the name and date of creation.  The clustering name actually denotes a sequence of clusterings generated over time that can be compared and contrasted.  As an example, we might cluster the Facebook friendship graph on a weekly basis.  The key-value service that provides clustering lookups is indexed by both the name and the date to identify the particular clusters.  

The universe is aware of both the clustering name and date upon creation.  To refresh a universe, all running experiments need to be halted within that universe to avoid changing the treatment assignments of running experiments.  We can then swap out the date of the clusters, while maintaining the clustering name.  This refresh provides continuity to experimenters who understand a particular universe as being appropriate for their experiments.
  
\section{Network experiment analysis}
\label{sec:analysis}

Before describing our estimation strategy, we describe how we identify treatment effects from our experimental design, focusing on estimands that can be straightforwardly estimated.  

\subsection{Basic estimand}
Our basic estimand from which we will construct other estimands of interest is 
\begin{eqnarray}
\mu(w, r) = \E[Y_u | W_u=w, R_u = r].  
\label{eq:basic_estimand}
\end{eqnarray}
This estimand answers the counterfactual question of \textit{what is the expected outcome of a random unit placed into condition $w$ under cluster ($r=1$) or unit ($r=0$) randomization under the experimental design}.  It is crucial to understand that there are two, and possibly three, aspects of randomness to the above expectation.  The first is the experimental design's allocation of the remaining population to conditions.  The second is the selection of the random unit in the population.  If we wish to refer to a specific unit $i$ in the population, instead of a randomly selected unit $u$, we will write $Y_i$. Finally, the outcome $Y_i$ for a given unit $i$ and $\textbf{W}$ could be considered as stochastic instead of deterministic~\cite{Pearl2009}.  We consider design-based inference where the only source of randomness is the experimental design.  In particular we assume 

\begin{assumption}  Every outcome $Y_i$, including $T_i$, is deterministic given treatment assignments $\textbf{W}$ for all units $i$.
\end{assumption}

A unit sampled into our experiment through unit-randomization is equivalent to a random unit in the population, so simple averages suffice to identify this estimand for $r = 0$.  A unit sampled via cluster-randomization is \textbf{not} a random unit in the population because this requires sampling clusters proportional to cluster size.  We can utilize importance sampling based on cluster size to adjust for this.  Letting $S_c$ represent the size of a cluster $c$ and $\E[]$ indicate the average over a randomly selected cluster $c$ in the population, we have that the basic estimand equals
\begin{align}
\mu(w, 1) &=\frac{\E [S_c \E[Y_u | W_u = w, R_u = 1, c_u = c]]}{\E[S_c]} \nonumber\\
&= \frac{\E[ \sum_i 1(c_i = c)Y_i | W_i = s, R_i = 1]}{\E[S_c]} \nonumber\\
&= \frac{\E[ Y_c | W_c = w]}{\E[S_c]}.
\label{eq:estimand_complex}
\end{align}
where in the final line we defined $Y_c$ as the sum over outcomes for all units in cluster $c$ and introduced $W_c \in \{0,1\}$ as the treatment assignment of cluster $c$, well-defined for a cluster-randomized cluster.  This final expression with informal notation is convenient because if we consider unit-randomized conditions as being drawn from a population of clusters with size $1$, then this estimand is valid for both the unit- and cluster-randomized conditions, although sampling clusters from different, albeit related, populations.  When the informal notation is non-ambiguous, we will use it.

\subsection{Other estimands}
From this basic estimand, we can construct various contrasts between conditions within the same population, including a difference-in-means estimand as
\begin{eqnarray}
\mu(1,1) - \mu(0,1) = \frac{\E[ Y_c | W_c = 1]}{\E[S_c]} - \frac{\E[ Y_c | W_c = 0]}{\E[S_c]},
\label{eq:diff_in_means}
\end{eqnarray}
and a ratio estimand as
\begin{eqnarray}
\frac{\mu(1,1)}{\mu(0,1)} - 1 = \frac{\E[ Y_c | W_c = 1]}{\E[ Y_c | W_c = 0]} - 1.
\label{eq:ratio_estimand}
\end{eqnarray}

We can also consider contrasts across unit- and cluster-randomized conditions which allow for testing if unit-level SUTVA holds~\cite{Saveski2017}.  If SUTVA is true, then in Eq.~\ref{eq:basic_estimand} the allocation of the remaining population to conditions is irrelevant.  Further, whether a unit was cluster-randomized or unit-randomized to treatment as indicated by $r$ is also irrelevant.  So SUTVA would imply that this difference-in-means
\begin{eqnarray}
\mu(w, 1) - \mu(w, 0) = \frac{\E[ Y_c | W_c = w]}{\E[S_c]} - \E[ Y_u | W_u = w, R_u = 0]
\label{eq:mixed_diff_in_means}
\end{eqnarray}  
is zero.  Beyond hypothesis testing, the magnitude of the difference can be informative of the magnitude of interference.

\subsection{Leveraging trigger logging}
\label{subsec:leveraging_trigger}
We can leverage trigger logging to make more precise inferences if we are willing to make assumptions about interference and trigger logging. We consider the following two assumptions:
\begin{assumption} 
(\textbf{SUTVA for triggering}) The outcome $T_i$, whether a unit is triggered or not, does not depend on their own treatment assignment and any other unit's treatment assignments, i.e., $P(T_i = 0 | \textbf{W}) = P(T_i = 0)$ for all $\textbf{W}$ and $i$.
\end{assumption}

\begin{assumption} 
(\textbf{Conditional SUTVA}) The outcome $Y_i$, conditioning on a lack of triggering, does not depend on their own treatment assignment and any other unit's treatment assignment, i.e., $\E[Y_i | \textbf{W}, T_i = 0] = \E[Y_i | T_i = 0]$ for all $\textbf{W}$ and $i$.
\end{assumption}

Applying all three assumptions, our basic estimand becomes
\begin{align}
\mu(w, r) = \bigr(&P(T_u = 1) \E[Y_u | W_u = w, R_u = r, T_u = 1] + \nonumber\\
&P(T_u = 0)\E[Y_u | T_u = 0]\bigr). \nonumber
\end{align}
The second term is not a function of $w$ or $r$ if these conditions hold, and therefore cancels out of contrasts across conditions.\footnote{The conditional SUTVA assumption is only needed for $r=1$ to be included in this statement.  The second term for $r=0$ would be $\E[Y_u | T_u = 0]$ without this assumption.}  Hence under these assumptions, we only need to estimate
\begin{eqnarray}
\E[Y_u | W_u = w, R_u = r, T_u = 1], \nonumber
\end{eqnarray}
because contrasts will not contain contributions from non-triggered units.\footnote{Strictly speaking, this is not true for the ratio estimand where the denominator should not condition on $T_u = 1$.  In this circumstance, we redefine our ratio estimand to include the conditioning.}  Filtering the experiment's population to triggered units can result in large gains in precision since these terms known to be zero are excluded from the contrasts.  However, those terms are only zero if the assumptions hold.

If SUTVA does not hold for triggering, then assuming SUTVA holds for other outcomes is unlikely.  We can test for SUTVA violations in triggering using Eq.~\ref{eq:mixed_diff_in_means} with triggering as the outcome, as well as Eq.~\ref{eq:diff_in_means} for both unit- and cluster-randomized contrasts.  In practice, we compare the average number of triggered units per triggered cluster of different cluster-randomized conditions.  They should be equal if SUTVA holds at the unit-level for triggering.

If this test passes, we assume SUTVA holds at the unit-level for triggering and test violations of conditional SUTVA for $Y$ given $T=0$ by comparing an estimate of $\E[Y_u | W_u = 1, R_u = 1, T_u = 0, \textbf{T}_{c_u} \neq \textbf{0}]$ to $\E[Y_u | W_u = 0, R_u = 1, T_u = 0, \textbf{T}_{c_u} \neq \textbf{0}]$ using Eq.~\ref{eq:ratio_estimand}. Strictly speaking this tests whether $\E[Y_i | \textbf{W}, T_i = 0, \textbf{T}_{c_i} \neq \textbf{0}]$ is inconsistent with $\E[Y_i | T_i = 0, \textbf{T}_{c_i} \neq \textbf{0}]$.

If these tests pass, we condition our analysis on just the population of triggered units.  If both tests do not pass, we drop both assumptions, focus on $r=1$, and fall back to assuming SUTVA at the cluster-level for triggering such that $P(\textbf{T}_c | \textbf{W}) = P(\textbf{T}_c | \textbf{W}_c)$ for any cluster $c$.  Combined with the discussion in Section~\ref{subsec:trigger}, this implies $P(\textbf{T}_c = \textbf{0} | \textbf{W}) = P(\textbf{T}_c = \textbf{0})$ and also that $\E[Y_u | W_u = w, R_u = 1, \textbf{T}_{c_u} = \textbf{0}] = \E[Y_u | \textbf{T}_{c_u} = \textbf{0}]$.  So we can filter the population of clusters to those with at least one triggered unit, keeping both triggered and non-triggered units in these clusters because the basic estimand for $r=1$ is then
\begin{align}
\mu(w, 1) = \bigr(&P(\textbf{T}_{c_u} \neq \textbf{0}) \E[Y_u | W_u = w, R_u = 1, \textbf{T}_{c_u} \neq \textbf{0}] + \nonumber\\
&P(\textbf{T}_{c_u} = \textbf{0}) \E[Y_u | \textbf{T}_{c_u} = \textbf{0}]\bigr). \nonumber
\end{align}
If partial interference for triggering does not hold, our estimates will contain additional bias.  We could test for this via exposure modeling applied to triggering, but many of our network experiments did not provide evidence of unit-level interference in triggering.

\subsection{Estimation}
\label{subsec:estimation}

Estimation for cluster-randomized trials generally proceeds via cluster-level summaries, mixed effect models, or generalized estimating equations~\cite{hayes2017cluster}.  For network experiments, we favor an approach that is simple to implement at scale and explain.

Our estimands can be estimated via sample averages.  In particular, we perform a conditional inference that views our experimental design as providing $k(w, r)$ samples of units or clusters with $W = w$ and $R = r$.\footnote{The rest of the population is assumed to be in control for the purpose of interference.}  Units in unit-randomized conditions ($r=0$) can be considered as clusters of size 1. If SUTVA for triggering and conditional SUTVA for $Y$ are deemed valid for the experiment, these clusters just consist of triggered units. Each cluster is accompanied by ($Y$, $W$, $R$, $X$, $S$), where $Y$ are experimental outcomes summed across the cluster, $W$ and $R$ define the treatment condition, $X$ are pre-treatment covariates summed across the cluster, and $S$ is the cluster size.

Let $\overline{A}(w, r)$ be the sample mean of $Y$, $X$, or $S$, or functions of these variables, 
across all clusters with $W=w$ and $R=r$ sampled into the experiment as follows. 
\begin{eqnarray}
\overline{A}(w,r) = \frac{1}{k(w, r)} \sum_c 1(W_c = w, R_c = r) A_c. \nonumber
\end{eqnarray}
For simplicity, we will hereafter omit the $(w, r)$ for any quantity that are defined on this group of clusters and only add this notation as needed for clarity.
An estimator for our estimand $\mu$ in Eq.~\ref{eq:basic_estimand} is then  
\begin{eqnarray}
\widehat{\mu} = \frac{\overline{Y}}{\overline{S}}. \nonumber
\end{eqnarray}
This estimator is asymptotically consistent, in that as the number of clusters $k$ goes to infinity, it equals our estimand~\cite{Middleton2015}.  For a finite number of clusters, the estimator is biased for $r=1$ because the average cluster size is random.  

To understand properties of this estimator, we apply the delta method (see~\cite{Deng2018} for a recent discussion) to second-order and derive the bias as 
\begin{eqnarray}
\E\left[\widehat{\mu} - \mu\right] \approx \frac{1}{\E[S]^2}\left(\mu Var(\overline{S}) - Cov(\overline{Y}, \overline{S})\right), \nonumber
\end{eqnarray}
where $\E[S]$ is the population mean of $S$. Note that this bias is zero if all clusters have the same size, a preference expressed in~\cite{Huan2015}.  This requirement is unnecessary though under the following:

\begin{assumption} Covariances of any two sample means, such as $Cov(\overline{Y}, \overline{S})$, 
under our experimental design are $O(1 / k)$.
\end{assumption}

Assuming SUTVA is sufficient but not necessary for this to happen, as including limited dependence between variables could still produce this asymptotic behavior~\cite{chen2004}.  Of course, we should keep in mind that with too strong of dependence, this asymptotic behavior may be incorrect.  With this caveat, the delta method shows the bias is $O(1/k)$ when $r=1$ and zero otherwise. 

Using the first-order term in the delta method expansion, we have
\begin{align}
Var(\widehat{\mu}) \approx \frac{1}{\E[S]^2}Var\left(\overline{Y} - \mu\overline{S}\right) \nonumber
\label{eq:delta_var}
\end{align}
and hence the variance is also $O(1 / k)$ under our assumption.  The root mean-squared error is then dominated by the standard deviation of order $O(1 / \sqrt{k})$.  A benefit of online experimentation is the experiments have a very large number of clusters, so we can consider any terms, like the bias, of order $O(1 / k)$ as irrelevant.  

At this point, one might consider the estimation largely complete.  We can compute $\widehat{\mu}$, and confidence intervals with additional assumptions, from the experimental data.  A difficulty with this approach is that the resulting confidence intervals can be very large, especially for contrasts between the unit-randomized and cluster-randomized conditions as typically $k(w, 1) \ll k(w, 0)$.  Variance reduction is required to produce reasonably-sized confidence intervals.

\subsection{Agnostic regression adjustment}
We achieve this variance reduction through regression adjustment.  We can define a collection of many sample means from the experimental data.  Unfortunately, we do not know any population means corresponding to these sample means, but we do have that, for some quantities (e.g. the pre-treatment covariates $X$), the contrast across conditions $(w, r)$ and $(w', r')$ is asymptotically vanishing
\begin{eqnarray}
\E[\widehat{\mu_{X}} - \widehat{\mu_{X}}'] \approx O(1/k) \approx 0, \nonumber
\end{eqnarray} 
where $\widehat{\mu_{X}}$ and $\widehat{\mu_{X}}'$ are simply $\frac{\overline{X}}{\overline{S}}$ for the conditions in the contrast. Let $\phi$ be a vector of such quantities (or linear combinations such as $\sum_{w, r} \beta_{wr} \widehat{\mu_{X}}(w,r)$ where $\beta_{wr}$ sums to zero), where $\E[\phi]$ is $O(1/k)$ and $Var(\phi)$ is $O(1/k)$.

Then our regression-adjusted estimator is defined as a function of a vector $\gamma$ with the same length as $\phi$
\begin{eqnarray}
\widehat{\mu_{\gamma}} = \widehat{\mu} - \gamma \phi. \nonumber
\end{eqnarray}
This quantity has bias $O(1/k)$ since both terms on the right side have a bias of $O(1/k)$. The variance of this adjusted estimator, considered as a function of $\gamma$, is
\begin{eqnarray}
Var(\widehat{\mu_{\gamma}}) = Var(\widehat{\mu}) - 2 \gamma Cov(\phi, \widehat{\mu}) + \gamma Var( \phi) \gamma \nonumber
\end{eqnarray}
with a minimum when
\begin{eqnarray}
\gamma^{*} = Var(\phi)^{-1} Cov(\phi, \widehat{\mu}). \nonumber
\end{eqnarray}  
The trick for regression adjustment is to estimate a $\widehat{\gamma}$ from experimental data (ideally $\gamma^{*}$).  Let $\widehat{\gamma}$ be a consistent estimator of some $\gamma$ that has bias $O(1/k)$ and variance $O(1/k)$.  We then have that
\begin{eqnarray}
\widehat{\mu_{\widehat{\gamma}}} - \widehat{\mu_{\gamma}} = - (\widehat{\gamma} - \gamma) \phi. \nonumber
\end{eqnarray}
This is the dot product of two terms of order $O(1/\sqrt{k})$ and (assuming a constant number of features) is therefore $O(1 / k)$.  The additional bias due to regression adjustment is therefore irrelevant, and we simply ignore that we estimated $\gamma$ such that
\begin{eqnarray}
Var(\widehat{\mu_{\widehat{\gamma}}}) \approx Var(\widehat{\mu}) - 2 \widehat{\gamma} Cov(\phi, \widehat{\mu}) + \widehat{\gamma} Var(\phi) \widehat{\gamma}. \nonumber
\end{eqnarray}
Estimating $\gamma^{*}$ directly appears difficult, so we apply the delta method to the covariances and variances in $\gamma^{*}$, and estimate both expanded expressions instead.  This will match $\gamma^{*}$ asymptotically, assuming the estimator is consistent and the delta method is valid.\footnote{Alternatively, one could estimate a reasonable $\gamma$ using weighted linear regression or even ordinary least squares.  The resulting expression will be different from the delta method estimator we utilize for $\gamma$ though.  Any reasonable estimating equation for $\gamma$ that has the appropriate asymptotic behavior should result in variance reduction.} 

What features should we choose for adjustment?  Like any prediction problem, features that are highly correlated with the outcome, yet not correlated with each other are useful.  Our standard approach is to utilize the pre-experiment value of the outcome of interest for the two conditions in a contrast.  This single-variable regression adjustment when applied to unit-randomized conditions has been referred to as CUPED~\cite{Deng2013, Xie2016}.  In specific, when we compute a contrast of $(w, r)$ to $(w',r')$, we let
\begin{eqnarray}
\phi = \widehat{\mu_{X}} - \widehat{\mu_X}' \nonumber
\end{eqnarray}
for a scalar pre-experiment $X$.

For completeness, our regression-adjusted difference-in-means estimand in Eq.~\ref{eq:diff_in_means} is then estimated by
\begin{eqnarray}
\widehat{\mu_{\widehat{\gamma}}} -\widehat{\mu_{\widehat{\gamma}}}' = \widehat{\mu} - \widehat{\mu}' - (\widehat{\gamma}+\widehat{\gamma}') \phi
\end{eqnarray}
and the ratio estimand in Eq.~\ref{eq:ratio_estimand} is estimated by
\begin{eqnarray}
\frac{\widehat{\mu_{\widehat{\gamma}}}}{\widehat{\mu_{\widehat{\gamma}}}'} - 1. \nonumber
\end{eqnarray}
For both estimators, we utilize the delta method to derive variance expressions, treating $\widehat{\gamma}$ terms as fixed non-random quantities.  

All of these tedious-to-derive variance expressions for the estimators (and $\widehat{\gamma}$) are in terms of covariances and variances of sample averages, such as $Var(\overline{Y})$, $Cov(\overline{S}, \overline{Y})$, $Cov(\overline{Y}, \overline{X})$, as well as population means.  To estimate these expressions, we can substitute in sample averages for the population means, but the covariances require additional assumptions.

\begin{assumption}  We assume an infinite population of clusters to ignore covariances across conditions.
\end{assumption}

Under this assumption, we can estimate sample mean covariances that do not include an outcome $Y$ via empirical covariances
\begin{align}
\widehat{Cov}(\overline{A}, \overline{B'}) = \begin{cases}
\frac{\widehat{Cov}(A, B')}{k} & \text{if}\ w = w', r= r'\\
0 & \text{otherwise}, \\
\end{cases}
\end{align}

Up to this point for the estimation procedure, we have only assumed asymptotic properties that can hold without SUTVA.

\begin{assumption}
For covariances including the outcomes $Y$, we assume the appropriate form of SUTVA holds (unit-level for $r=0$ and cluster-level for $r=1$) such that the above expression is also valid for estimating covariances with $Y$.
\end{assumption}
Therefore unlike our point estimates, our asymptotic standard errors and associated confidence intervals rely upon the appropriate form of SUTVA.

\subsection{Brief note on model-based alternatives}
\label{sec:alternatives}

Agnostic design-based analysis is of course not the only way to analyze experiments.  A class of alternatives are model-based approaches, which broadly speaking, fit a model to experimental results, and use the fitted model to extrapolate the global average treatment effect $\tau$.

For example, regression modeling, like that described by~\cite{chin2019}, would attempt to estimate $\mu(\textbf{w}, \textbf{x}) = \E[Y_u | \textbf{W}=\textbf{w}, \textbf{X}=\textbf{x}]$ from the results of an experiment, by assuming that the dependence of $\mu$ on $\textbf{w}$ and observed covariates $\textbf{x}$ can be summarized through features, such as the number of nearby units in treatment to the random unit $u$, and covariates of $u$ and their nearby units.  If this model including covariates and features based on treatment conditions was correct (and was learned from data), evaluating it at $\textbf{w}=\textbf{1}$ and $\textbf{w}=\textbf{0}$ could be used to estimate the global average treatment effect $\tau$.  Unlike the agnostic regression adjustment, which we just used for improving precision, this approach requires trusting the model is well-specified enough to be useful.

A tremendous advantage of model-based analysis is that it produces an actual estimate of the global average treatment effect $\tau$, as opposed to $\tau_{cluster}$.  However, we are not aware of a satisfactory solution to fitting models, performing model validation and criticism across many experiments in many domains at scale.  
Considering that experts can have difficulty constructing and validating such causal models, asking engineers to do it themselves is a non-starter.  Given this state, we view agnostic regression adjustment as a safe default that trades bias for robustness, scale, and simplicity.  This may not be the best approach for any particular intervention, but is less likely to be dramatically wrong than extrapolation from an ill-formulated model.  That being said, nothing in our framework precludes experts from building models on top of data generated by a network experiment.  Modeling is an analysis choice that can be complementary to cluster-randomized designs.

\section{Cluster design} \label{sec:clusters}
A good clustering of experiment units is important for network experiments.  An ideal clustering will include all interference within clusters so that there is no interference between clusters, which removes the estimand bias in $\tau_{cluster}$.  To be more specific, we consider the setting of graph-cluster randomization where we assume a domain expert has created a relevant graph $G$ where each unit in the population is a vertex and edges, possibly weighted, represent a hypothetical interference pattern among the units.  The creation of the graph requires strong domain knowledge, but we do not require that this graph precisely represent interference.  A higher quality graph will just lead to higher quality clusters.

Let purity denote the fraction of edges within clusters.  We assume a clustering with higher purity has less bias.  A naive approach that achieves 100\% purity, and hence captures all interference, is grouping all units into a giant single cluster.  This is obviously unacceptable for a cluster-randomized experiment and so purity is just one aspect of cluster quality.  In addition, an experiment should have enough statistical power to detect treatment effects.  A single cluster including all units has no power, and a clustering that puts every unit in its own cluster, equivalent to unit randomization, leads to good power but the worst purity (zero). The tradeoff between purity and power is a \textit{bias-variance tradeoff}: higher purity leads to less bias while more statistical power requires smaller clusters. 

We consider two prototypical clustering algorithms:  Louvain community detection (Louvain)~\cite{blondel2008fast} and recursive balanced partitioning (BP)~\cite{kabiljo2017social}.  We chose these algorithms for discussion because they have extremely scalable implementations.  Crucially, these algorithms can produce a very large number of clusters from graphs representing large populations, which was shown essential for sufficiently low variance estimation in Section~\ref{subsec:estimation}.  Moreover, the two algorithms are also emblematic of clustering approaches that produce imbalanced and balanced cluster sizes.  In particular, Louvain generally produces very imbalanced clusters with a heavy-tailed cluster-size distribution, whereas BP generates balanced clusters.  

We find that \textit{imbalanced graph clusters are typically superior in terms of the bias-variance tradeoff for graph-cluster randomization}.  Other clustering algorithms that scale to many clusters may have different tradeoffs and can be properly evaluated via the tools we provide.  While an extensive comparison across such algorithms is outside the scope of the paper, we suggest that imbalanced clustering algorithms, like Louvain, should always be considered.  To demonstrate this, we describe our evaluation procedure.      

\subsection{Evaluating the bias-variance tradeoff}
To evaluate a clustering's statistical power for experiments, we run synthetic Monte-Carlo AA tests assuming partial interference holds for the given clustering and apply the estimation procedure described in Section~\ref{subsec:estimation}.  If similar experiments have been run before in the past, we can reuse trigger logging from the past experiments, otherwise we assume a random fraction of the population is triggered on each simulation.  For outcome $Y$, we consider the primary metric of interest and let $X$ be pre-synthetic experiment values of the same metric to enable cluster-based regression adjustment.  

After validating that statistical properties of the estimator are reasonable (i.e. coverage, CI width, etc.)\footnote{The estimation procedure can perform poorly if outlier clusters contain a non-trivial fraction of all units in the population.}, we produce an estimate of the minimal detectable effect (MDE), which can be considered as a simple transformation of the estimator standard deviation.  A larger standard deviation means a larger MDE. We represent our bias-variance tradeoff as a MDE-purity tradeoff, plotting one against the other.  An ideal clustering would have high purity and low MDE.

\subsection{Representative results}
We describe results for a particular clustering, but we emphasize that these results are qualitatively similar to our general experience designing clusters for graph-cluster randomization. 

\subsubsection{Clustering generation}

We applied BP and Louvain with various parameters on a graph consisting of about 90 billion edges and about 2 billion vertices. Louvain optimizes modularity~\cite{newman2004finding}, a quantity related to purity. Louvain includes a resolution parameter, and in our implementation, smaller resolution leads to smaller clusters.  We consider $\{0.001, 0.0001, 0.00001\}$ for the resolution parameter.  Our Louvain implementation repeatedly iterates on a clustering across iterations, and we consider evaluating clusters produced at iterations from 2 to 5.  A single run of BP also provides multiple clusterings through recursively generating a binary tree.  Each level of the binary tree corresponds to a partition of units, where level $k$ contains $2^k$ clusters.  We looked at BP clusters produced by levels $17, 18, 19,$ and $20$ corresponding to $2^{17}$, $2^{18}$, $2^{19}$, $2^{20}$ clusters. 

\subsubsection{Cluster size distribution}
\begin{figure}[!ht]
	\centering
	\includegraphics[width=\columnwidth]{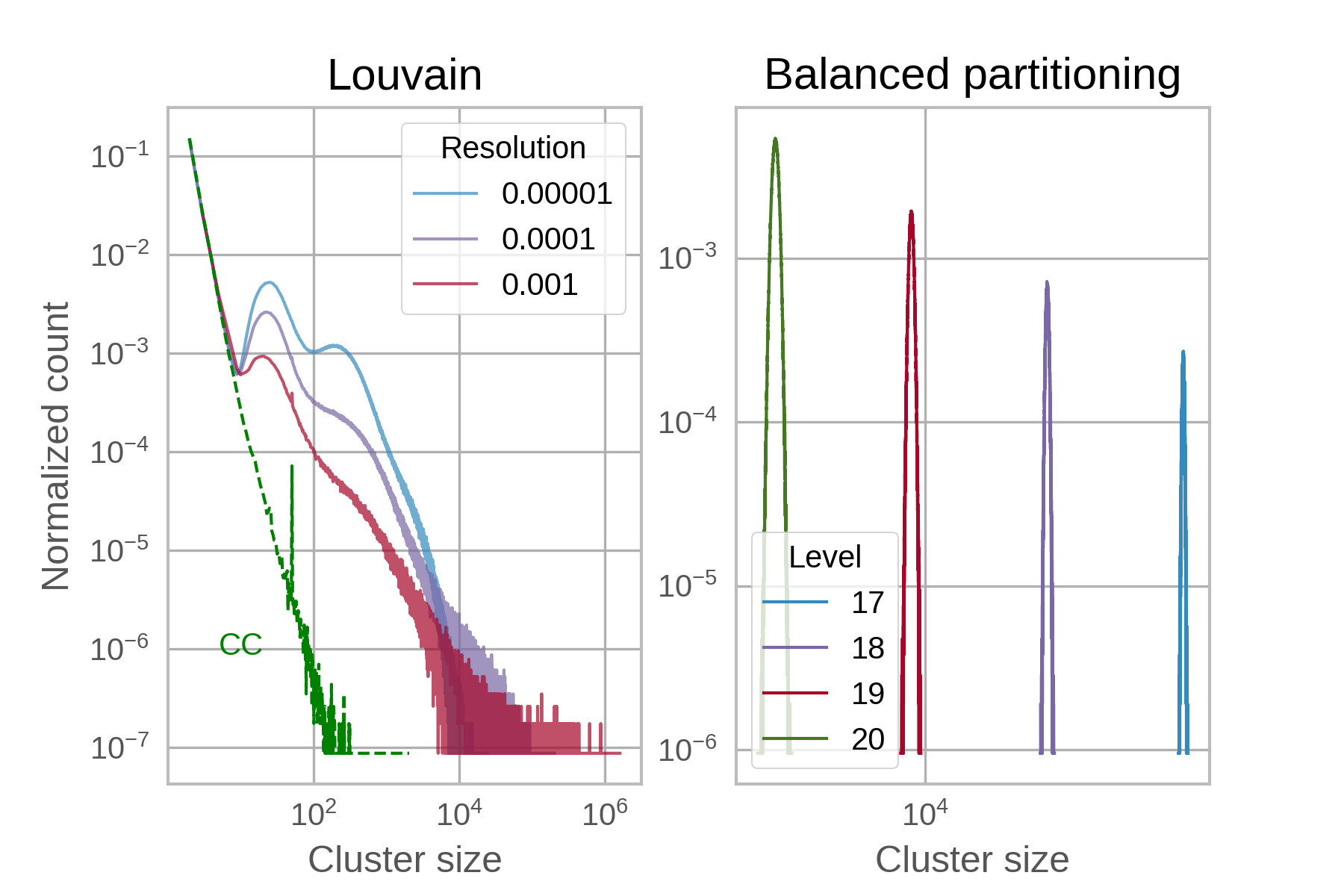}
	\caption{Cluster size distribution from (left) Louvain with different resolution and (right) recursive BP at different splitting level. Y-axis is the number of clusters at a specific cluster size, normalized by the total number of clusters.
	}
	\label{fig:bp_lv_size_dist}
\end{figure}

Fig.~\ref{fig:bp_lv_size_dist} shows the cluster size distributions from Louvain (left, iteration = 3, similar patterns for other iterations) and from recursive BP (right). In general, Louvain follows a heavy-tailed distribution, while BP results in a much more balanced size distribution.  We observe smaller resolution leads to smaller clusters in the tail but more clusters in the middle size range.  On the left (small) side of the distribution, it follows very closely with the distribution from small connected components (CC) shown by the green dashed line, as expected from modularity maximization. The peak around 50 in the CC distribution is an artifact of filtering applied to the graph.  From the right plot, BP maintains a tight distribution of cluster sizes across levels, with a shifting mean cluster size. 

\subsubsection{MDE-purity tradeoff}
\begin{figure}[!ht]
	\centering
	\includegraphics[width=\columnwidth]{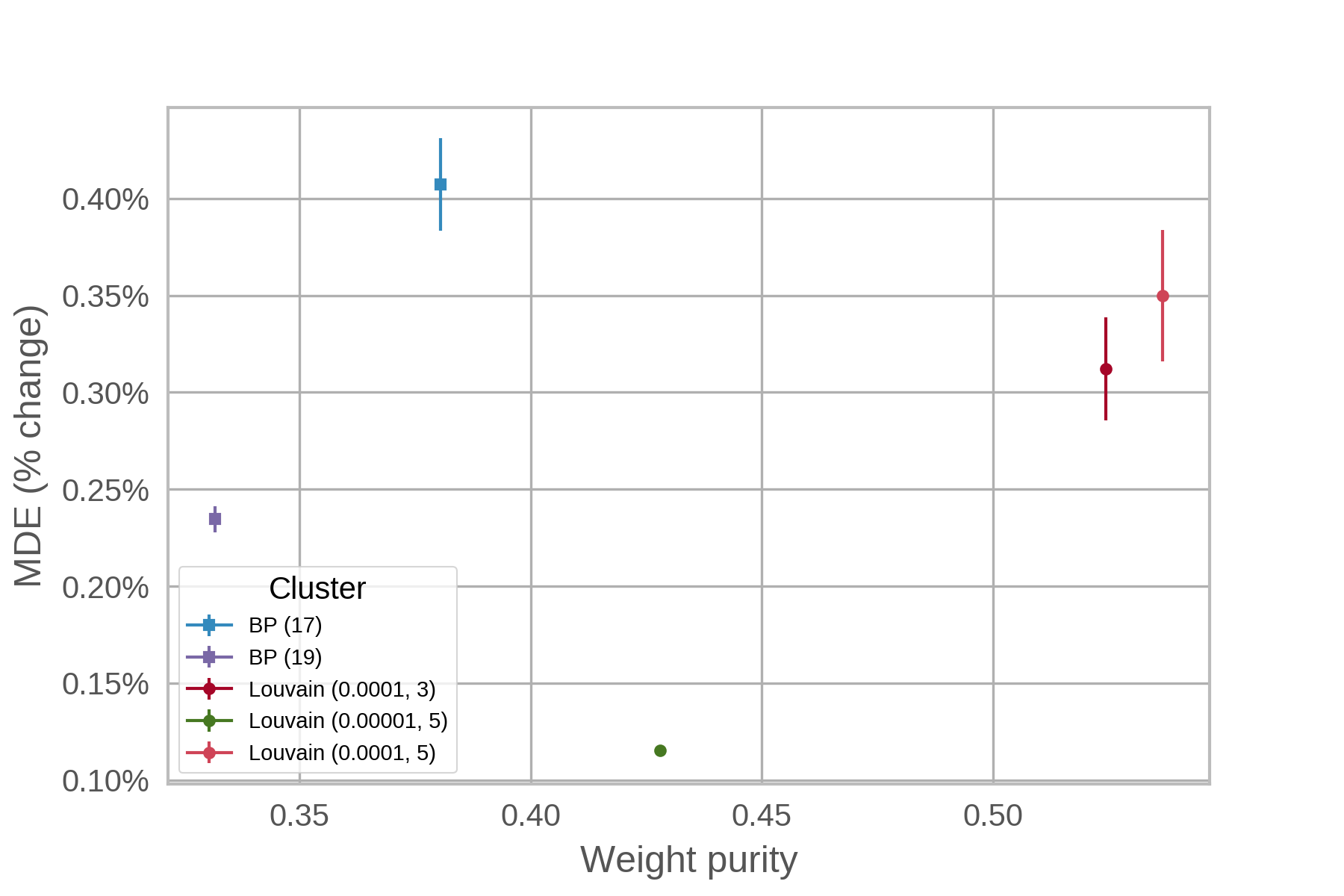}
	\caption{MDE vs purity for Louvain and recursive BP with different parameters: BP ($x$) is BP with level $x$, and Louvain ($a$, $b$) is Louvain with resolution $a$ and iteration $b$.
	}
	\label{fig:bp_lv_mde}
\end{figure}

For each clustering, we estimated the MDE at a given power (95\%). The tradeoff between MDE and purity is shown in Fig.~\ref{fig:bp_lv_mde}: for either Louvain or BP, a higher purity is often associated with a larger MDE.  We observe the two algorithms perform very differently in terms of purity and MDE. In particular, Louvain generates imbalanced clusterings with much higher purity but also comparable MDE to BP, which provides evidence against claims that balanced clusters should be preferred for graph-cluster randomization.

\section{Network experiment case studies}
\label{sec:examples}

We now describe two network experiments in detail, one using graph-cluster randomization and another based on geographic clusterings.

\subsection{Stories experiment}
\begin{figure}[!ht]
	\centering
	\includegraphics[width=0.3\columnwidth]{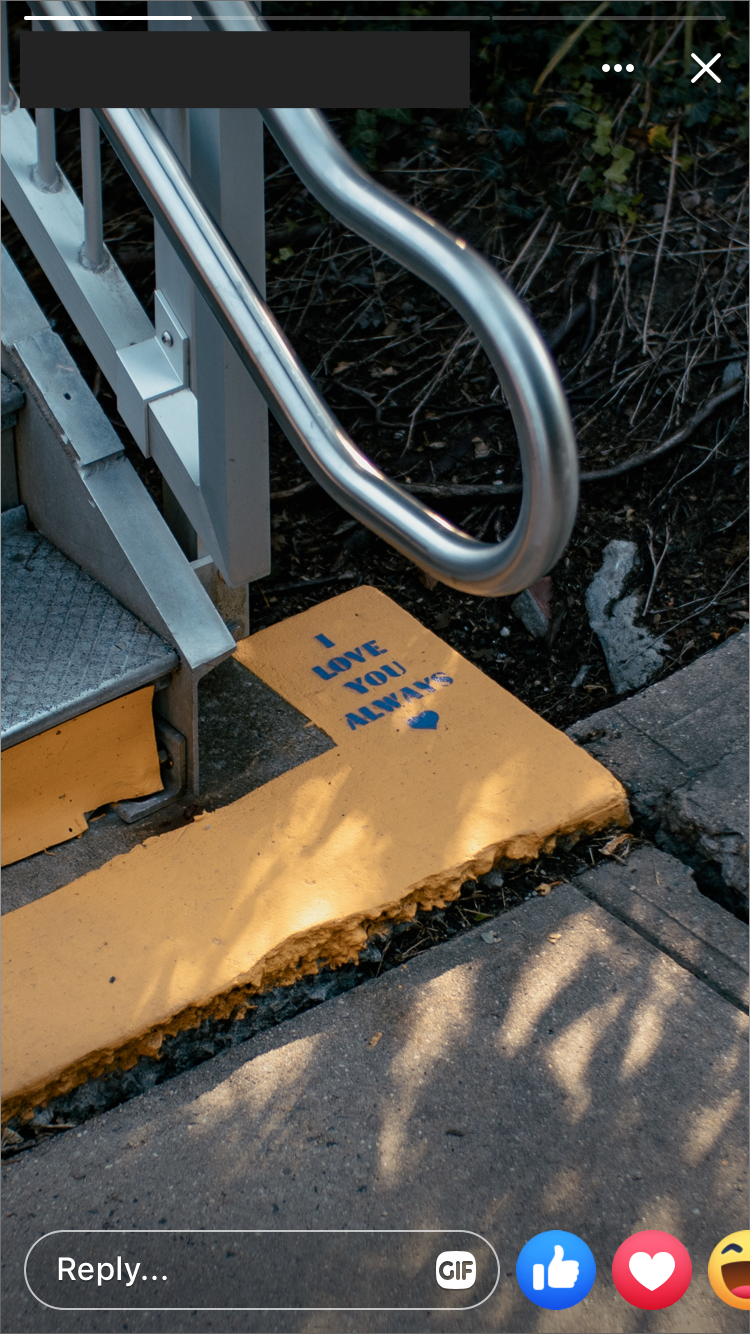}
	\includegraphics[width=0.3\columnwidth]{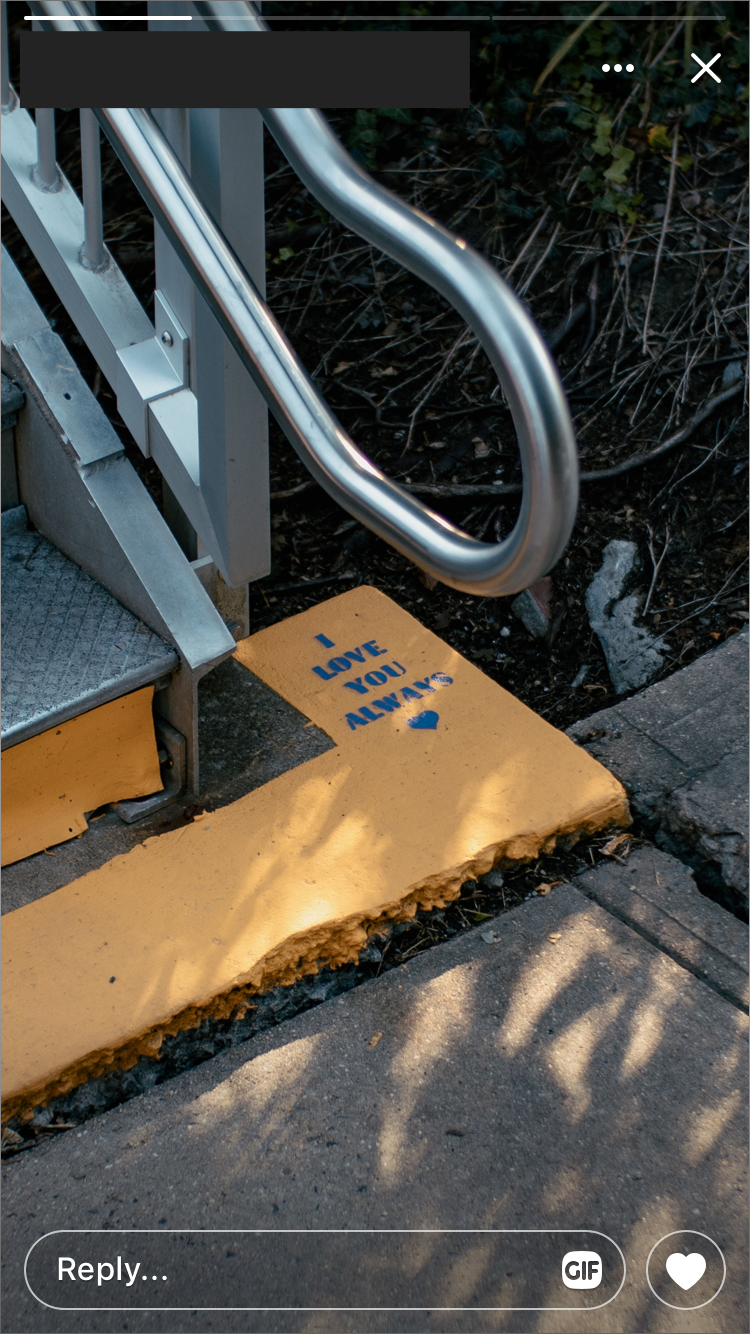}
	\includegraphics[width=0.3\columnwidth]{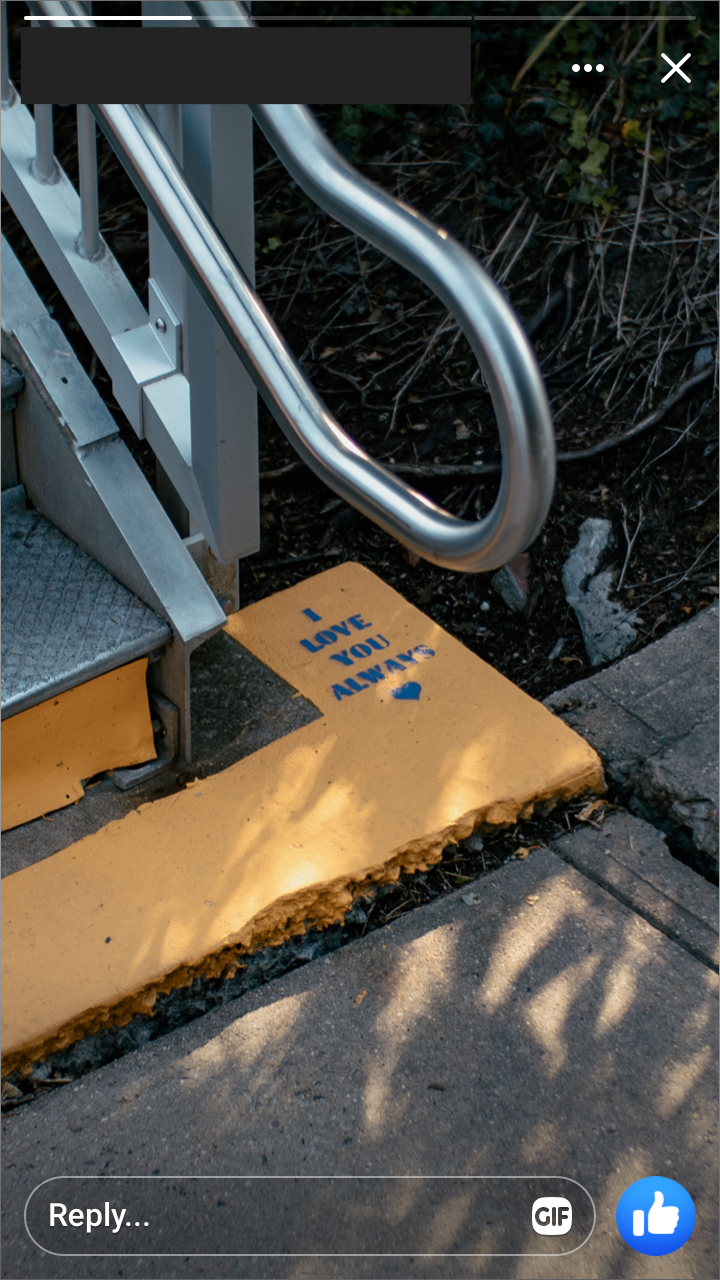}
	\caption{Test and control in Stories experiment. (left) Control group: all emojis can be seen by horizontally scrolling the bottom bar; (middle) Test group for iOS: static transparent heart; (right) Test group for Android: static blue thumb-up. To access all emojis in the test group we need to click on either the heart or the thumb-up.}
	\label{fig:feed_story_intervention}
\end{figure}

The Facebook Stories Viewer Experience team conducted a two-weeks-long mixed network experiment for Android and iOS users on a new reply design: in the control group, when users view a Facebook story they have a horizontal scroll bar (left in Fig.~\ref{fig:feed_story_intervention}) for all emojis and in the test group users have either a blue thumb-up (Android, right in Fig.~\ref{fig:feed_story_intervention}) or a transparent heart (iOS, middle Fig.~\ref{fig:feed_story_intervention}) next to the text reply box. All emojis will then pop up after clicking on the thumb-up or heart. 

This new feature potentially alters a user's experience as both a story \textit{viewer} and as a story \textit{creator}. Previous user-randomized experiments indicated that the new feature increases text replies sent to stories, but decreases the usage and send rate of emojis. Such user-randomized tests did not detect a treatment effect on story creators, since they received feedback from viewers in both the test and control groups.  To estimate indirect treatment effects of this feature change on users who create stories, we ran a mixed experiment expected to reduce the interference between test and control viewers for story creators and estimate the presence and magnitude of network effects.  In this network experiment, users were grouped into clusters using Louvain community detection algorithm based on historical interaction data for Facebook Stories. 

\subsubsection{Hypothesis}
\begin{figure}[!ht]
	\centering
	\includegraphics[width=\columnwidth]{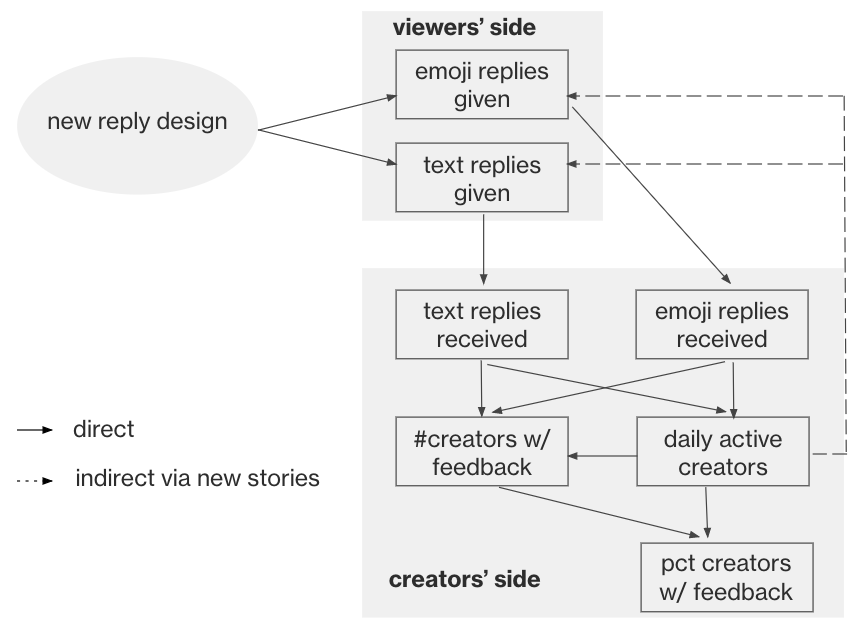}
	\caption{Hypothetical causal graph of metrics for Stories experiment.}
	\label{fig:feed_story_causal_graph}
\end{figure}

The metrics of interests include: emoji replies from story viewers, text replies from viewers, emoji replies received by story creators, text replies received by story creators, daily active story creators, number of story creators with at least one feedback and percentage of creators receiving at least one feedback.  Fig.~\ref{fig:feed_story_causal_graph} shows a hypothetical causal graph of the metrics: the new reply design has direct effects on viewer responses to stories, which affects the story creators' replies received, and number of story creators with feedback.  Creator production of new stories then has an indirect effect on a viewer's future reply metrics.

The hypothesis is that the new experience would increase text replies and decrease lightweight replies from viewers. With the mixed experiment, the primary interest is the downstream treatment effects on story creator metrics and understanding how cluster-randomized estimates compare to user-randomized estimates. Since cluster-randomization reduces test-control interference on story creators from viewer replies, we expect to measure larger differences between the user- and cluster-side estimates for story creator metrics than story viewer metrics. 

\begin{table*}
	\centering
	\caption{Estimated ATE for different contrasts in the Stories mixed experiment on triggered subpopulation}
	\label{tab:user_cluster_ate}
	\begin{tabular}{lcccc} 
		\hline \hline
		Metric & cluster test - cluster ctrl   & user test - user ctrl & cluster test - user test   & cluster ctrl - user ctrl   \\ 
		\hline
		\textbf{Story Viewer Metrics} & & & & \\
		emoji replies given      &  $-23.84\% \pm 0.69\% $                                   & $-23.62\% \pm 0.16\% $ &$ -0.19\% \pm 0.36\% $&$ 0.20\% \pm 0.35\% $   \\ 

		text replies given    &  $11.06\% \pm 0.56\% $  & $11.76\% \pm 0.17\% $     &$ -0.57\% \pm 0.36\%$&$ -0.06\% \pm 0.34\%  $ \\ \\
		\textbf{Story Creator Metrics} & & & & \\

		emoji replies received & $-6.37\% \pm 0.87\%$ & $0.73\% \pm 2.44\%$ &$ -5.90\% \pm 2.38\% $&$ 1.40\% \pm 0.73\%$ \\ 

		text replies received & $2.01\% \pm 0.56\%  $   & $-0.13\% \pm 0.26\% $     &$ 1.69\% \pm 0.44\% $ & $ -0.52\% \pm 0.44\%$ \\ 

		percent of creators with feedback  & $-0.91\% \pm 0.07\% $&$-0.05\% \pm 0.04\%$  &$ -0.78\% \pm 0.06\%$ & $ 0.10\% \pm 0.05\%$\\ 

		number of creators with feedback      & $-1.16\% \pm 0.21\%   $& $-0.21\% \pm 0.08\%  $&$ -0.96\% \pm 0.16\% $&$ 0.00\% \pm 0.16\%$  \\ 

		daily active creators & $-0.25\% \pm 0.18\%$ &$-0.17\% \pm 0.07\%$  &$ -0.18\% \pm 0.14\% $&$ -0.09\% \pm 0.14\%  $ \\
		\hline
	\end{tabular}
\end{table*}

\subsubsection{Results}
Point estimates and 95\% confidence intervals of the ratio estimators for cluster-randomized conditions are shown in Table~\ref{tab:user_cluster_ate} in the column "cluster test - cluster ctrl". The results for text and emoji replies are consistent with our expectations: emoji replies decreased by 23.8\% while the text replies increased by 11.1\%, resulting in a large drop in replies sent to stories overall. The movement in the replies received by story creators are consistent with the replies sent by viewers, though the effect sixes are smaller: -6.4\% vs -23.8\% for emoji replies and 2\% vs 11\% for text replies received and sent, respectively.  Since story creators still receive replies from test and control viewers (the cluster purity is about 40\%), this difference is expected. The cluster-randomized experiment also shows a 0.25\% drop in daily active creators and a 1.2\% drop in the number of active creators with feedback, resulting in in a 0.9\% drop in the percent of story creators with feedback. 

As expected, story creator metrics in the user-side are not significant and smaller than in the cluster-side, due to large test-control interference from viewers, which provides evidence on the effectiveness of cluster-randomization to reduce interference. To quantify the difference in user- and cluster-side estimates, we directly compare the cluster-randomized and user-randomized test groups in the last two columns of Table~\ref{tab:user_cluster_ate}. The estimates of the cluster test group for all story creator metrics are significantly different from those of the user test group. By design, producers in the cluster test group receive replies more often from viewers within the same cluster, hence receiving less emoji replies and more text replies, resulting in less replies overall, thus decreasing daily active creators. The same applies for the number of story creators with feedback, yet compounded with the effects on the viewer reply metrics.

In this experiment the viewer experience is similar in both the cluster- and user-randomized conditions. Thus, the difference between cluster and user conditions are much smaller (-0.19\% for emoji replies given and -0.57\% for text replies given) than the overall treatment effect (about -24\% for emoji replies and +10\% for text replies on both sides). These small differences are possibly due to the indirect effect from the drop in daily active creators.

These estimates are consistent with our hypothetical causal graph in Figure \ref{fig:feed_story_causal_graph}. The large differences in story creator metrics relative to the overall treatment effects provides evidence that network experiments capture more interference than previous user-randomized designs and lead to more reasonable estimates.\footnote{As a check, we computed the pre-experiment contrasts between conditions and found no significant pre-experiment differences. We also checked the experiment differences between cluster-control and user-control, which are not significant in most cases. Between cluster-control and user-control there is a small significant difference in the percent of creators with feedback and in the creator reply metrics, likely due to test-control interference from viewers.}

\subsubsection{Methodology variations}
\begin{figure}[h]
	\centering
	\includegraphics[width=\columnwidth]{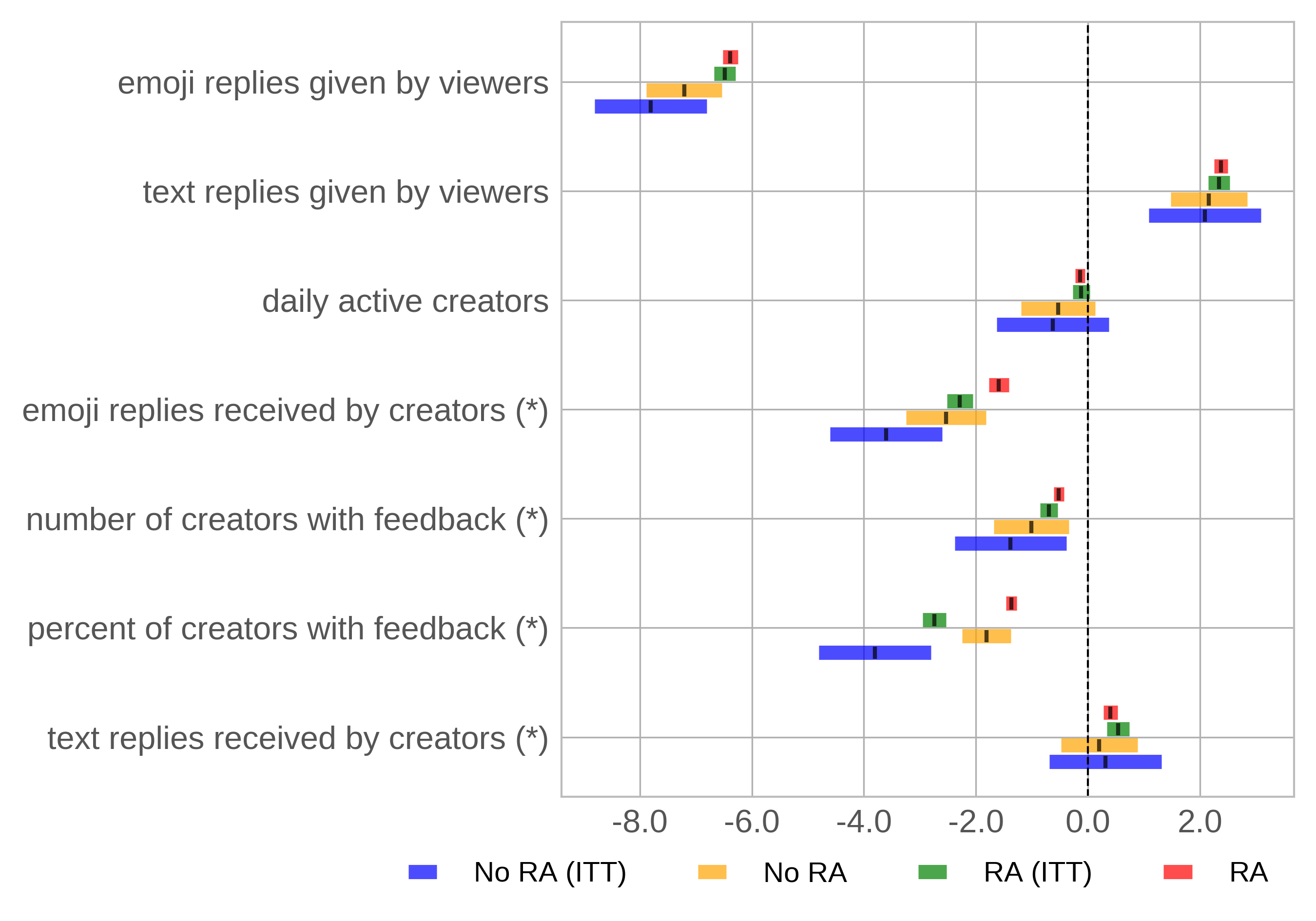}
	\caption{Comparison of ATE estimates with scaled 95\%  confidence intervals computed on triggered users and triggered clusters (ITT), with and without regression adjustment (RA) for cluster test versus cluster control in the Stories experiment. The story creator metrics labeled with (*) do not pass the conditional SUTVA check.}
	\label{fig:variance_reduction}
\end{figure}

So far, we have only presented results on the sub-population of triggered users using regression adjustment.  In Figure~\ref{fig:variance_reduction}, we show how results for the cluster test and cluster control comparison change if we perform an Intent-To-Treat (ITT) analysis on the triggered clusters as described in Section~\ref{subsec:leveraging_trigger}, instead of triggered users, and if we do not use regression adjustment.  Each metric row was re-scaled such that the ITT estimate without regression adjustment has a confidence interval (CI) width of one.  The figure shows how point estimates and CI's change based on the method used.

We clearly see that regression adjustment (RA) provides substantial precision gains.  Conditioning on the triggered users provides additional gains, although of smaller magnitude.  Looking closely at the comparison between RA (ITT) and RA, we do see significant differences for some metrics.  Indeed, the metrics labeled by a star are ones for which RA (ITT) may be more accurate.  For this experiment, we concluded that SUTVA for triggering was valid, but our testing for conditional SUTVA for $Y$ depended on the metric of interest.  Indeed, most story creator side metrics, labeled with a star, failed this conditional SUTVA test, and we see evidence of this in the different estimates provided by RA and RA (ITT).

We note though that while the ITT estimates may be mildly different on the starred creator metrics, the qualitative interpretation of the results on the triggered clusters remains consistent with the results on the triggered users presented in the previous subsection.  

\subsection{Commuting Zones experiment}
Given the nature of Facebook, a natural way to cluster individuals might be through friendship networks or online engagements between content producers and consumers, as in the previous subsection.  Network experiments can be valuable for any sort of clustering though. The nature of the product, or the treatment being tested (and the source of interference) can help determine the relevant clustering over which we should randomize our experiment.

Sometimes, a relevant network is defined by geographical location. At Facebook this might be particularly the case for products like Jobs on Facebook (JoF). For some products, such as JoF, geographical clusters may be especially appropriate as individuals are likely to interact with employers closer to their own physical location.

\begin{figure}[h]
	\centering
	\includegraphics[width=0.35\columnwidth]{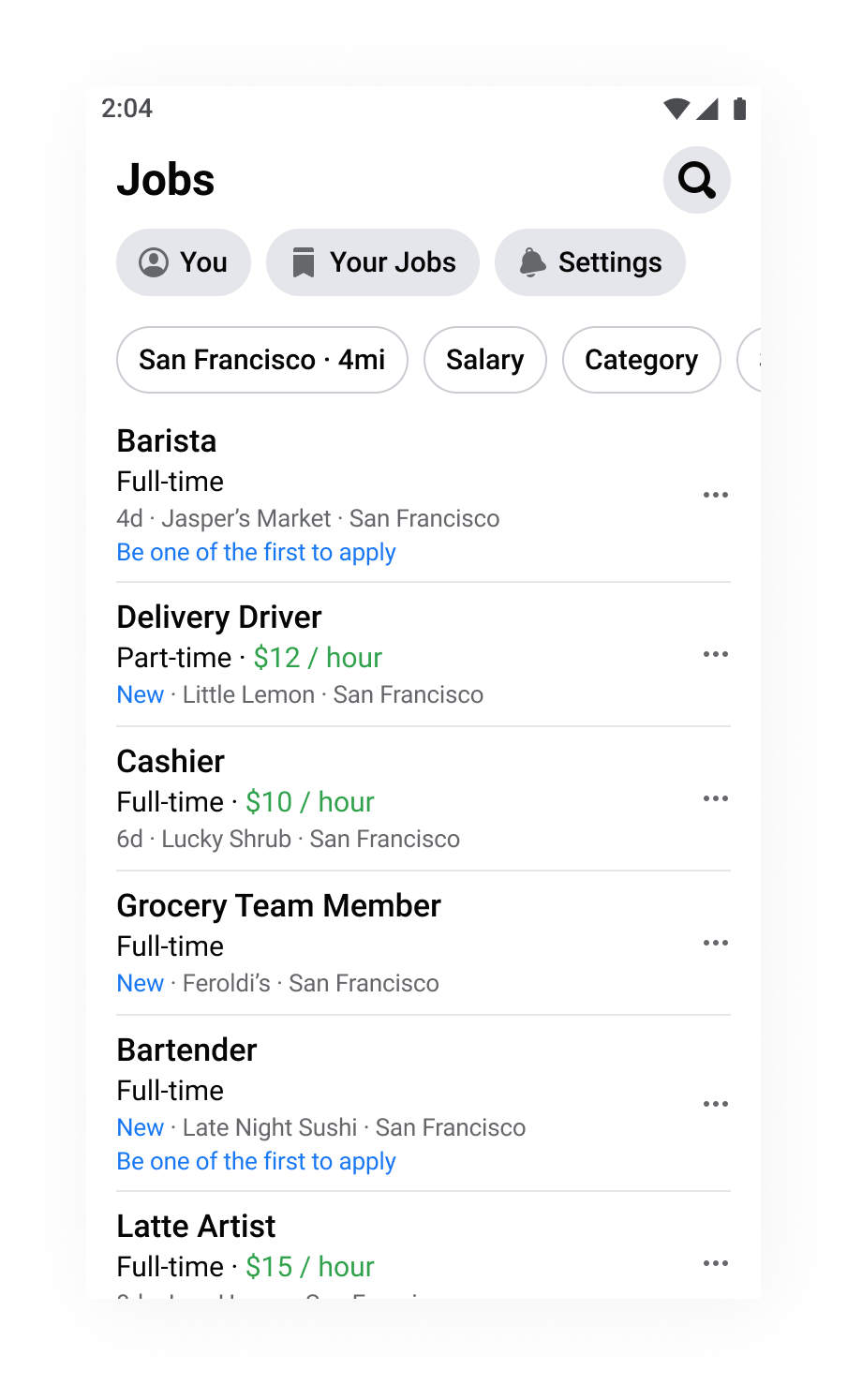}
	\caption{Jobs on Facebook}
	\label{fig:jobs_browser}
\end{figure}

We describe an example of such an experiment for JoF.  JoF connects hiring businesses to FB users by enabling employers to list jobs to which users can apply.  In this example, the platform seeks to connect users to employers who may be most likely to hire them and seeks to provide a good experience to both users and employers. Given that this is a two-sided market, changes that affect user behavior may also have an impact on employer behavior, and isolating these total effects is difficult in a user-randomized experiment. Figure \ref{fig:jobs_browser} shows an example of the Facebook Jobs Browser, which provides a list of employment opportunities in a users general area. 

The JoF team wanted to understand the overall ecosystem effects of up-ranking jobs that had few previous applications at the time a user was on the Jobs Browser. Connecting users to jobs with fewer applications can help users and employers by connecting users to employers who are most likely to hire them, and by improving the experience for employers with few previous engagements (who might get the highest marginal utility from an application). This may lead employers to post more jobs, allowing the platform to provide more opportunities to future job seekers. 

In a user test, the team boosted jobs with few previous applications in the browser and tracked applications sent on the platform as well as applications sent to jobs that had no previous applications at the time of submission. This second metric served as a proxy for the impact on creators, since it is difficult to measure this impact directly in a user-randomized experiment. One issue with this metric is interference occurs between the treatment and control groups. By boosting jobs without previous applications, we increase the visibility of these jobs in the treatment group which leads more users to apply to these jobs. When users in the control group apply, these jobs now have one application or more, but some of these users would have applied to the job anyway.  This interference causes a user test to overstate the treatment effect of our boost on this metric of applications to jobs with no previous applications.

Since users search for jobs in their general location, the team ran this experiment at the Facebook commuting zone level, available through Facebook's Data for Good program, to account for interference between users in the treatment and control groups \cite{cz2020}. An example of these clusters are shown below in Figure \ref{fig:fbcz_example}.

Since jobs on the platform are associated with a location, each job can also be assigned to a commuting zone. With the commuting zone randomization, the team could directly measure the effect of this change on job listings and employers for the first time. More specifically, we could test whether providing applications to jobs with few applications leads employers to post another job. To understand these effects, we focus on jobs that were posted before the test started, as well as employers who had posted a job before the test started. Including jobs that were created during the experiment would pose a problem if employers do in fact post more as a result of receiving more applications. In this analysis, jobs and pages are considered to be exposed to the experiment as soon as the first user in their commuting zone is logged as triggered.

\begin{figure}[h]
	\centering
	\includegraphics[width=\columnwidth]{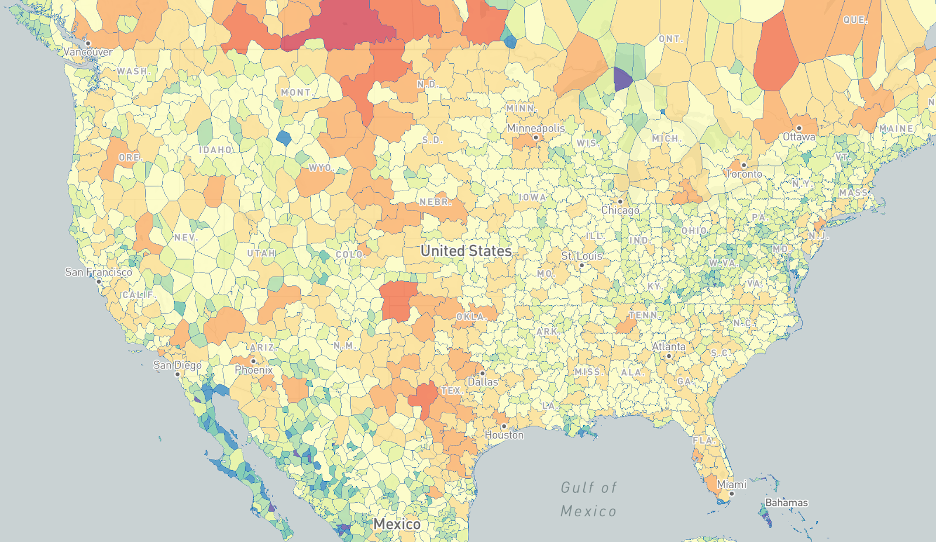}
	\caption{Facebook Commuting Zones in North America}
	\label{fig:fbcz_example}
\end{figure}

\begin{table*}[ht!]
\caption{Commuting Zones experiment results}
\label{tab:cz_test_results}
\begin{tabular}{lcc} 
\\[-1.8ex]\hline 
\hline \\[-1.8ex] 
Metric & Unit of Randomization & Estimated Effect (with 95\% CI)  \\ 
\hline \\[-1.8ex] 
Applications to jobs with no previous applications & user & 71.841\% $\pm$ 5.087\% \\
Applications to jobs with no previous applications & cluster & 49.652\% $\pm$ 17.817\% \\
Probability a job receives an application & cluster & 14.069\% $\pm$ 11.198\% \\
Probability an employer posted a new job & cluster & 16.959\% $\pm$ 15.731\% \\
\hline \\[-1.8ex] 
\end{tabular} 
\end{table*}

\subsubsection{Results}

The team ran a mixed experiment, as described above, where trigger logging was useful since a small fraction of the population is looking for new jobs at any given time. Table \ref{tab:cz_test_results} below summarizes the results of this test from a 2.5 week period. In the user-side results, the metric of interest (applications to jobs with no previous applications) increased by 71.8\%. These results were consistent with a user test that was run before the mixed test.  The commuting zone mixed experiment, however, showed that the user-side treatment effects were upwardly biased. Instead of increasing applications to jobs with no previous applications by 71.8\%, the cluster-randomized estimate was a 49.7\% increase, a difference that is statistically significant at the 5\% level.  In the absence of these results the team might have inferred they were increasing applications to these jobs more than was truly the case.

The cluster-randomized standard errors for this metric are much larger (17.8\%) than the user-randomized standard errors (5.1\%). While expected due to the randomization over clusters, some of this difference is also due to the relatively small number of commuting zones. More specifically, there are fewer than 20,000 commuting zones across the world. In general, regression adjustment is extremely insightful for these kinds of experiments, where we have seen that adjustment can reduce the size of the standard errors in commuting zone experiments by over 30\%.

The team also learned that by increasing applications to jobs with few overall applications by 49.7\%, they were able to increase the probability that a job receives an application by 14.1\% and the probability that an employer posted another job increased by 17.0\%. Both of these results were statistically significant at the 5\% level.

By randomizing this experiment at the commuting zone (geographical) level, the team confirmed a number of hypotheses. First, user-randomization leads to significant bias in the metric of applications to jobs with zero previous applications. Second, changes to the user experience that increase this metric do in fact cause employers to post more jobs on the platform. Understanding the interactions between applicants and employers in a two-sided marketplace is important for the health of such a marketplace, and network experiments can enable understanding these interactions better than user-randomized tests. 

\section{Discussion}

We have introduced a practical framework for designing, implementing, and analyzing network experiments at scale. Our implementation of network experimentation accommodates mixed experiments, cluster updates, and the need to support multiple concurrent experiments.  The simple analysis procedure we present results in substantial variance reduction by leveraging trigger logging as well as our novel cluster-based regression adjusted estimator. We also introduce a procedure for evaluating clusters, which indicated bias-variance tradeoffs are in favor of imbalanced clusters, and allows researchers to evaluate the tradeoffs for any clustering method they would like to explore.

The two mixed experiments, one leveraging graph-cluster randomization and another geographic clusters, demonstrate the flexibility and value of network experiments.  In both cases, with the help of large precision gains from regression adjustment, we estimated significant interference effects compatible with our hypotheses, with clear evidence of differences between cluster- and user-randomized conditions.  These effects had been hidden in prior user-randomized experiments testing the same changes, indicating that network experiment results are not merely differences in magnitude, but can provide differences in interpretation. In the latter experiment we were also able to measure effects that were only possible with cluster-randomization. 

These two experiments were chosen by the authors for the purpose of illustration.  Not all tests benefit from cluster-randomization or show significant network effects in a mixed experiment, especially if interference is weak or non-existent relative to the direct treatment effects.  Our expectation is that many product changes at Facebook, and similar online services, may not require a network experiment for accurate estimates of the global average treatment effect.  Under weak interference, the bias-variance tradeoff favors user-randomized designs.  

That said, we believe that interference is often ignored for the purpose of convenience and not due to prior knowledge.  Through our framework experimenters can learn whether interference is relevant in their domain by allowing them to easily run a network experiment and find out.  On the flip side of ignoring interference, we also have observed engineers claim that interference would have led to improved results from those seen in user-randomized experiments.  We can now test these claims and avoid hypothetical discussions about the direction and magnitude of interference.  

As we have discussed, our design-based results are only unbiased under partial interference, and we view network experiments as an approach that trades bias for robustness, simplicity, scale, and convenience.  One compelling avenue for future research is to consider alternative estimation procedures for experimental data generated from network experiments that carefully applies additional modeling assumptions.

\bibliographystyle{ACM-Reference-Format}
\bibliography{netexp} 

\end{document}